\documentclass{aa} 
\usepackage{graphicx}
\usepackage{txfonts}
\begin{document}

\title{Can giant planets form by gravitational fragmentation of discs?}

\subtitle{Radiative hydrodynamic simulations of the inner disc region ($\stackrel{<}{_\sim}40$ AU) }

\titlerunning{Can giant planets form by gravitational fragmentation of discs?}
\author{D. Stamatellos \& A. P. Whitworth}

\institute{School of Physics and Astronomy, Cardiff University,
           5 The Parade, Cardiff CF24 3AA, Wales, UK\\
           \email{D.Stamatellos@astro.cf.ac.uk}\\
		\email{A.Whitworth@astro.cf.ac.uk} 
           }
 
\date{Received September, 2007; accepted }

\abstract
{Gravitational fragmentation has been proposed as a mechanism for the formation of giant planets in close orbits around solar-type stars. However, it is debatable whether this mechanism can function in the inner regions ($R\stackrel{<}{_\sim}40$ AU) of real discs.}
{We investigate the thermodynamics of the inner regions of discs and their propensity to fragment.}
{We use a newly developed method for treating the energy equation and the equation of state, which accounts for radiative transfer effects in SPH simulations of circumstellar discs. The different chemical and internal states of hydrogen and the properties of dust at different densities and temperatures (ice coated dust grains at low temperatures, ice melting, dust sublimation) are all taken into account by the new method.}
{We present radiative hydrodynamic simulations of the inner regions of massive circumstellar discs and examine two cases: (i) a disc irradiated by a cool background radiation field ($T_{_{\rm BGR}}\sim 10\,{\rm K}$) and (ii) a disc heated by radiation from its central star ($T_{_{\rm BGR}}\!\sim\! 1200{\rm K}[R/{\rm AU}]^{-1}$). In neither case does the disc fragment: in the former because it cannot cool fast enough and in the latter because it is not gravitationally unstable. Our results (a) corroborate previous numerical results using different treatments for the hydrodynamics and the radiative transfer, and (b) confirm our own earlier  analytic predictions.} 
{Disc fragmentation is unlikely to be able to produce giant planets around 
solar-type stars at radii $\stackrel{<}{_\sim} 40$~AU.}


\maketitle

\section{Introduction}
A large number of exoplanets ($\sim 265$) have been discovered around nearby stars in the last 12 years, mainly with the radial velocity method  (Udry \& Santos 2007; {\it Extrasolar planets Encyclopaedia} at {\tt http://exoplanet.eu}). Most of these exoplanets have properties that are different from those in the Solar System. In particular, giant planets are found on very close orbits of apparently random eccentricity. 

Two main theories have been proposed for the formation of giant planets, (i) core accretion, and (ii) gravitational fragmentation. In the core accretion scenario giant planets form by coagulation of planetesimals (e.g. Safronov 1969; Goldreich \& Ward 1973; Pollack et al. 1996). Once a solid body of around $10~{\rm M}_{_{\bigoplus}}$ is reached, it quickly accretes a large gaseous envelope. One of the main problems with this theory is that the timescale for planet formation is long. The theory predicts that planets can form within a few million years, but observations suggest that  circumstellar discs may not be that long lived (Haisch, Lada \& Lada 2001; Cieza et al. 2007). In the gravitational fragmentation scenario, giant planets form by gravitational instability in massive discs (e.g. Kuiper 1951; Cameron 1978; Boss 1997; Durisen et. al 2007). The main advantage of the gravitational fragmentation theory is that planets condense out on a dynamical timescale, i.e. $\la 10^3\,{\rm years}$. 

There are two conditions that must be fulfilled  for discs to fragment gravitationally. (i) The disc must be gravitationally unstable, i.e. massive enough so that gravity can overcome thermal pressure and centrifugal support (Toomre 1964),
\begin{equation}
Q(R)\;=\;\frac{c(R)\,\kappa(R)}{\pi\,G\,\Sigma(R)}\;\la\;1\,,
\end{equation}
where $c$ is the isothermal sound speed, $\kappa$ is the epicyclic frequency, $\Sigma$ is the surface density, and $R$ is the distance from the axis of rotation. (ii) A proto-fragment must be able to cool fast enough  for the energy delivered by compression to be radiated away. Theory and simulations (Gammie 2001; Johnson \& Gammie 2003; Rice et al. 2003, 2005;  Mayer et al. 2004; Mejia et al. 2005) indicate that the cooling must happen on a dynamical time-scale, 
\begin{equation}
t_{_{\rm COOL}}\; <\;{\cal C}(\gamma)\,t_{_{\rm ORB}}\,,\hspace{1.0cm}0.5\la{\cal C}(\gamma)\la 2.0\,,
\end{equation}
where $t_{_{\rm ORB}}$ is the local orbital period, and $\gamma$  the adiabatic exponent. 

However, it is uncertain whether {\it real discs} actually satisfy the above conditions. Boss (2004)  and Mayer et al. (2007) suggest that convection  provides the necessary cooling, whereas Johnson \& Gammie (2003),  Boley et al. (2006) and Nelson (2006) find that the cooling is too slow, and hence fragmentation is not possible. The latter point of view is supported by analytic studies which indicate that convection cannot provide the required cooling  (Whitworth et al. 2007; Rafikov 2007), and that discs cannot cool fast enough, at least close ($R\stackrel{<}{_\sim}50$~AU) to the central star (Rafikov 2005; Matzner \& Levin 2005; Whitworth \& Stamatellos 2006).

One reason why hydrodynamic simulations have produced contradictory results concerning whether disc fragmentation can produce giant planets close to the central star is the different treatments of radiative transfer used. We have recently developed an efficient scheme to capture the thermal and  radiative effects when protostellar gas fragments (see Section 2 and Stamatellos et al. 2007). Thus, we are able  to perform radiative SPH simulations of the inner disc regions and to include the effects of the equation of state consistently, i.e. by solving the relevant Saha equations and taking into account the resulting continuous and differentiable changes in the mean molecular weight and the internal energy. Moreover our radiative scheme allows us to include irradiation of the disc by a background radiation field. Our simulations suggest that it is very difficult for planets to form by gravitational instability in the {\it inner} regions of a massive circumstellar disc.\footnote{For simulations of gravitational fragmentation in the {\it outer} regions of a massive circumstellar disc, the reader is referred to Stamatellos, Hubber \& Whitworth (2007).} 

The structure of the paper is as follows. In Section 2 we describe our radiative hydrodynamic code. In Section 3 we define the initial conditions. In Section 4 we present and discuss the simulations. In Section 5 we summarise the results and their implications for the possibility of forming giant planets close to a star by gravitational fragmentation.
 
\section{Numerical method}

For the hydrodynamics we use the SPH code {\sc dragon} (Goodwin et al. 2004), which invokes an octal tree (to compute gravity and find neighbours), adaptive smoothing lengths, multiple particle timesteps, and a second-order Runge-Kutta integration scheme. The code uses time-dependent viscosity with parameters $\alpha^\star=0.1$, $\beta=2\alpha$  (Morris \& Monaghan 1997) and a Balsara switch (Balsara 1995), so as to reduce artificial shear viscosity (Artymowicz \& Lubow 1994; Lodato \& Rice 2004; Rice, Lodato, \& Armitage 2005).

The energy equation is treated with the method of Stamatellos et al. (2007). This method uses the density and the gravitational potential of each SPH particle to define a pseudo-cloud around each particle, through which the particle cools and heats. The mass-weighted mean column-density $\bar{{\Sigma}}_i$, and the Rosseland-mean opacity  $\bar{\kappa}_{_{\rm R}}(\rho_i,T_i)$, averaged over every possible position of the particle inside its pseudo-cloud, are then used to calculate the net radiative heating rate for the particle, according to
\begin{equation} 
\label{eq:radcool}
\left. \frac{du_i}{dt} \right|_{_{\rm RAD}} =
\frac{\, 4\,\sigma_{_{\rm SB}}\, (T_{_{\rm BGR}}^4-T_i^4)}{\bar{{\Sigma}}_i^2\,\bar{\kappa}_{_{\rm R}}(\rho_i,T_i)+{\kappa_{_{\rm P}}}^{-1}(\rho_i,T_i)}\,;
\end{equation}
here, the positive term on the right hand side represents heating by the background radiation field, and ensures that the gas and dust cannot cool radiatively below the background radiation temperature $T_{_{\rm BGR}}$. $\;{\kappa_{_{\rm P}}}(\rho_i,T_i)$ is the Planck-mean opacity, and $\sigma_{_{\rm SB}}$ the Stefan-Boltzmann constant. 

The method takes into account compressional heating, viscous heating, radiative heating by the background, and radiative cooling. It performs well, in both the optically thin and optically thick regimes, and has been extensively tested (Stamatellos et al. 2007). In particular it reproduces the detailed 3D results of Masunaga \& Inutsuka (2000), Boss \& Bodenheimer (1979), Boss \& Myhill (1992), Whitehouse \& Bate (2006),  and also the analytic test of Spiegel (1957). It is efficient and easy to implement in particle- and grid-based codes. Because the diffusion approximation is applied here globally, the method does not capture in detail the exchange of heat between neighbouring fluid elements. Our simulations also have insufficient resolution to capture convection.

The gas is assumed to be a mixture of hydrogen and helium. We use an equation of state (Black \& Bodenheimer 1975; Masunaga et al. 1998; Boley et al. 2007a) that accounts (i) for the rotational and vibrational degrees of freedom of molecular hydrogen, and (ii) for the different chemical states of hydrogen and helium. However, we note that in the simulations presented here the temperature never becomes hot enough for significant dissociation of H$_2$, and consequently the mean molecular weight is approximately constant. We assume that ortho- and para-hydrogen are in equilibrium.

For the dust and gas opacity we use the parameterization  by Bell \& Lin (1994), $\kappa(\rho,T)=\kappa_0\ \rho^a\ T^b\,$, where $\kappa_0$, $a$, $b$ are constants that depend on the species and the physical processes contributing to the opacity at each $\rho$ and $T$. The opacity changes due to ice mantle melting, the sublimation of dust, molecular and H$^-$ contributions,  are all taken into account. 

\section{Disc initial conditions}

We simulate a $0.07~{\rm M}_{\sun}$ disc around a $0.5 ~{\rm M}_{\sun}$ star. This is a relatively massive disc, but such discs have been observed, for example in the Orion Nebula cluster (Eisner \& Carpenter 2006). The disc initially extends from 2 to 40~AU. Its initial surface density and temperature are 
\begin{eqnarray}
\label{eq:sdensprofile}
\Sigma(R)&=&\Sigma_0\,\left(\frac{R}{\rm AU}\right)^{-1/2}\,,\\
\label{eq:tempprofile}
T(R)&=&\left[T_0^2 \left(\frac{R}{AU}\right)^{-2}+T_{\infty}^2\right]^{1/2}\,,
\end{eqnarray}
where $\Sigma_0=1.8\times 10^{-4}\,{\rm M}_\odot\,{\rm AU}^{-2}$ and $T_0=1200\,{\rm K}$ are the surface density and temperature at $R\simeq 1\,{\rm AU}$, and $T_{\infty}=10$~K is the asymptotic temperature far from the central star. These are typical disc profiles and have been chosen so as to match approximately the initial density and temperature profiles used in the simulations of Boley et al. (2006) and Cai et al. (2007). We assume that the initial  disc temperature is independent of distance from the disc midplane.

To calculate the initial disc thickness, $z_0(R)$, we balance the vertical gravitational acceleration due to the star and the underlying disc, against the hydrostatic acceleration, 
\begin{eqnarray}
\frac{G M_{\star}}{R^2}\frac{z_0(R)}{R}+\pi G\Sigma(R)\;\simeq\;\frac{c^2(R)}{z_0(R)},\hspace{3.3cm}&&\\
\label{eq:disc.thickness}
z_0(R)\;\simeq\;-\frac{\pi\Sigma(R)R^3}{2 M_{\star}}+\left[{\left(\frac{\pi\Sigma(R)R^3}{2 M_{\star}}\right)^2+\frac{R^3}{G M_{\star}}c^2(R)}\right]^{1/2}\,,&&
\end{eqnarray}
where $c(R)$ is the local isothermal sound speed.

For the initial vertical structure of the disc,  we adopt a sinusoidal profile 
\begin{equation}
\rho (R,z)=\rho(R,0)\cos\left[\frac{\pi z}{2 z_0(R)}\right],\hspace{0.5cm}|z|<z_0(R)\,.
\end{equation}
This profile resembles a Gaussian profile (e.g. Frank, King \& Raine 1992). Setting $\Sigma (R)=\int_{-z_0(R)}^{z_0(R)}{ \rho (R,z) dz}\,$, we obtain
\begin{equation}
\label{eq:disk.dens}
\rho (R,z)=\frac{\pi\,\Sigma_{0}}{4\,R^{1/2}\,z_0(R)}\,\cos\left[\frac{\pi\,z}{2\,z_0(R)}\right],\hspace{0.5cm}|z|<z_0(R)\,.
\end{equation}

The choice of the initial density and temperature profiles is not critical, since the disc quickly relaxes to a quasi-equilibrium state.

\section{Simulations}

We perform two simulations, one with a uniform blackbody background radiation field having temperature $T_{\rm BGR}=10~K$ (cf. Boley et al. 2006; Cai et al. 2007), and one which takes account of radiation from the central star ($T_{\rm BGR}\sim R^{-1}$).  We use $2\times10^5$ SPH particles to represent the disc. This provides more than enough resolution to capture fragmentation, since the Jeans condition (e.g. Bate \& Burkert 1997) is easily satisfied everywhere (by more than a factor of one hundred in mass).  The central star is represented by a sink with radius 2~AU. An SPH particle is accreted by the sink if it is within the sink radius and is bound to the sink. The central star is allowed to move.

\begin{figure}
\centering{
\includegraphics[height=9.cm,angle=-90]{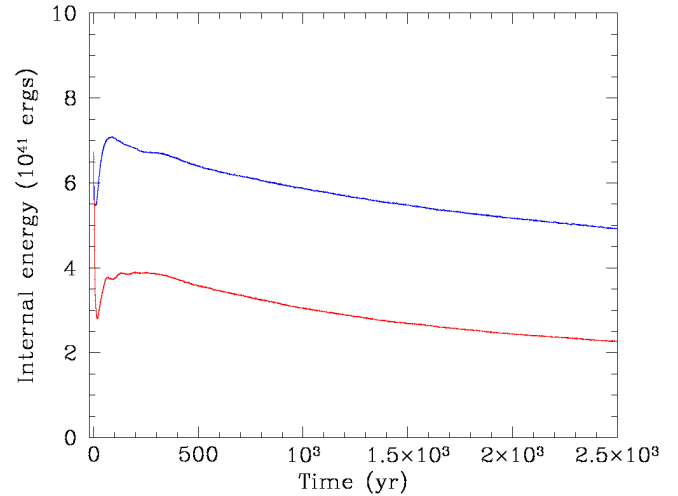}}
\caption{Internal energy (i) for a disc irradiated by a uniform $10\,{\rm K}$ background radiation field (bottom, red line) and (ii) for a disc irradiated by its central star ($T_{_{\rm BGR}}(R)\sim R^{-1}$; top, blue line).}
\label{fig:discene}
\end{figure}

In both cases the disc relaxes from the initial conditions to a quasi-steady state.  In Fig.~\ref{fig:discene} we plot the evolution of the disc internal energy with time for both cases. With a uniform background radiation field at $T\sim 10\,{\rm K}$, the disc relaxes to a lower-temperature quasi-steady state (bottom line) than in the case where radiation from the central star is taken into account (top line). Otherwise the evolution of the thermal energy appears similar in the two cases. In the following subsections we describe each simulation in detail.

\begin{figure}
\centering{
\includegraphics[height=9cm,angle=-90]{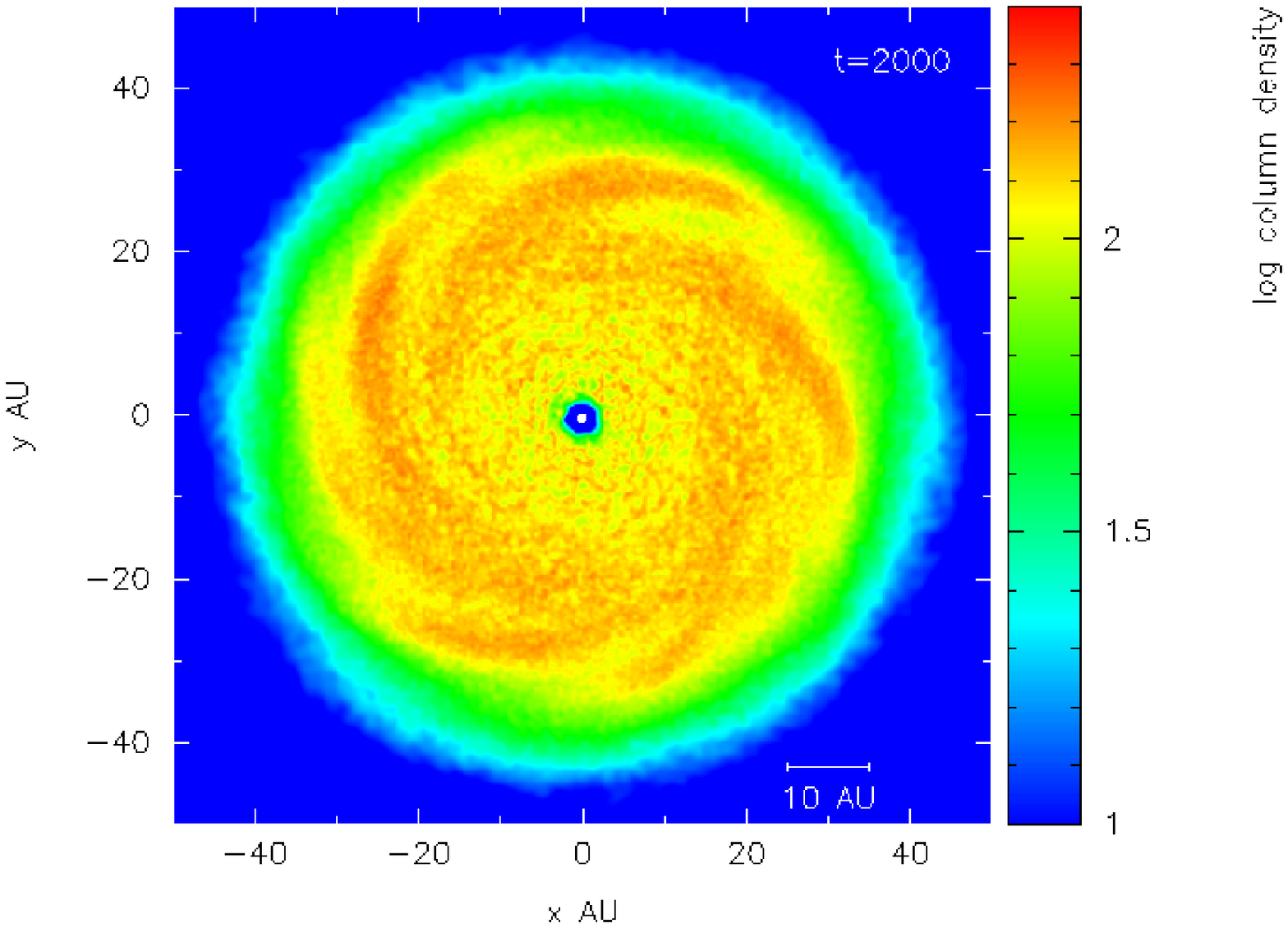}
\includegraphics[height=8.3cm,angle=-90]{dens.xz.4b.ps}}
\caption{Logarithmic column density in ${\rm g}\,{\rm cm}^{-2}$, projected on the $xy$-plane (top), and volume density in ${\rm g}\,{\rm cm}^{-3}$ on the $xz$-plane (bottom) for a disc irradiated by a $10\,{\rm K}$ background radiation field.}
\label{fig:dur4b.dens}
\centering{
\includegraphics[height=9cm,angle=-90]{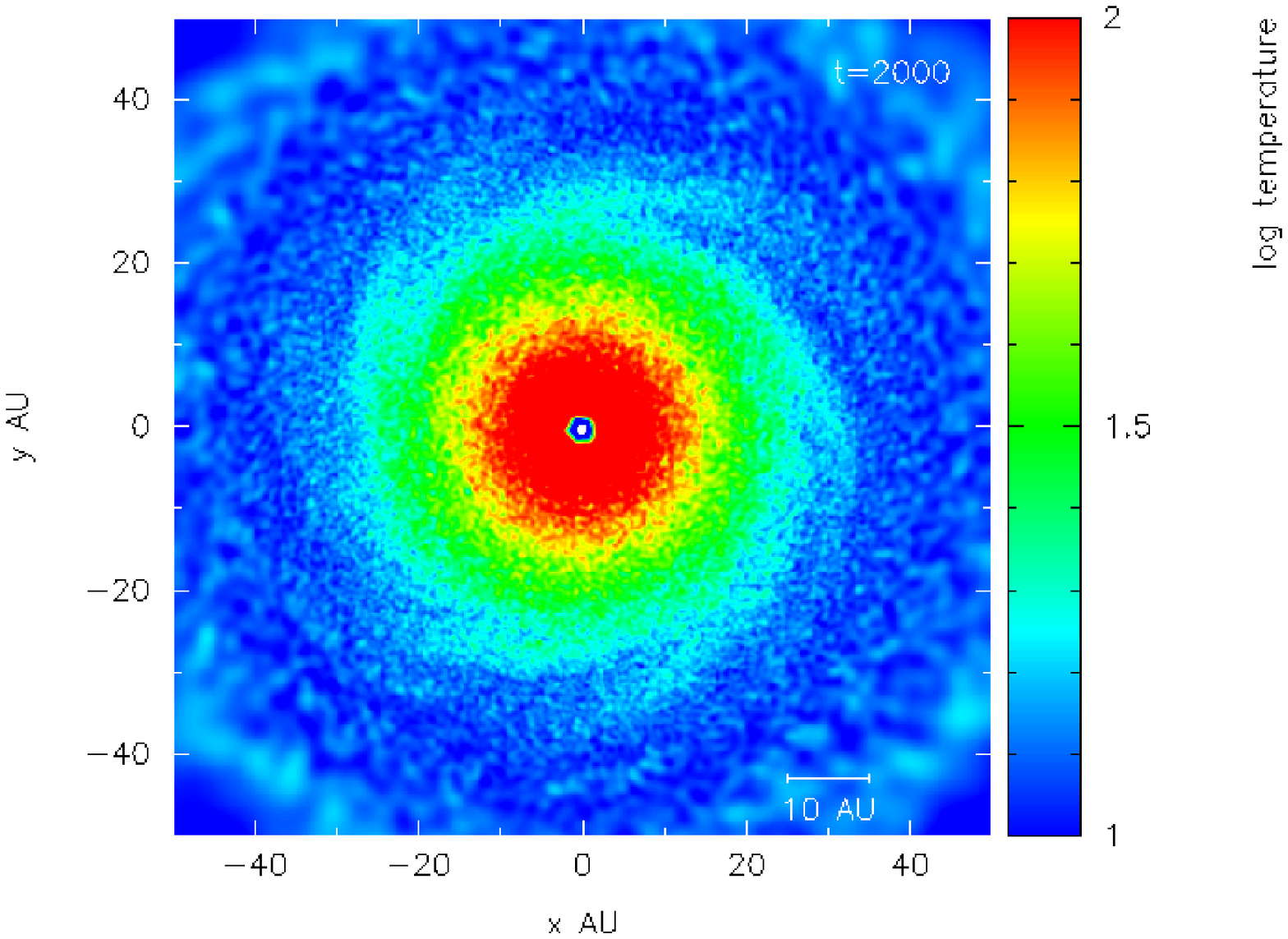}
\includegraphics[height=8.3cm,angle=-90]{temp.xz.4b.ps}}
\caption{Logarithmic temperature on the $xy$-plane (top), and on the $xz$-plane (bottom) for a disc irradiated by a $10\,{\rm K}$ background radiation field.}
\label{fig:dur4b.temp}
\end{figure}

\begin{figure}
\centering{
\includegraphics[height=9cm,angle=-90]{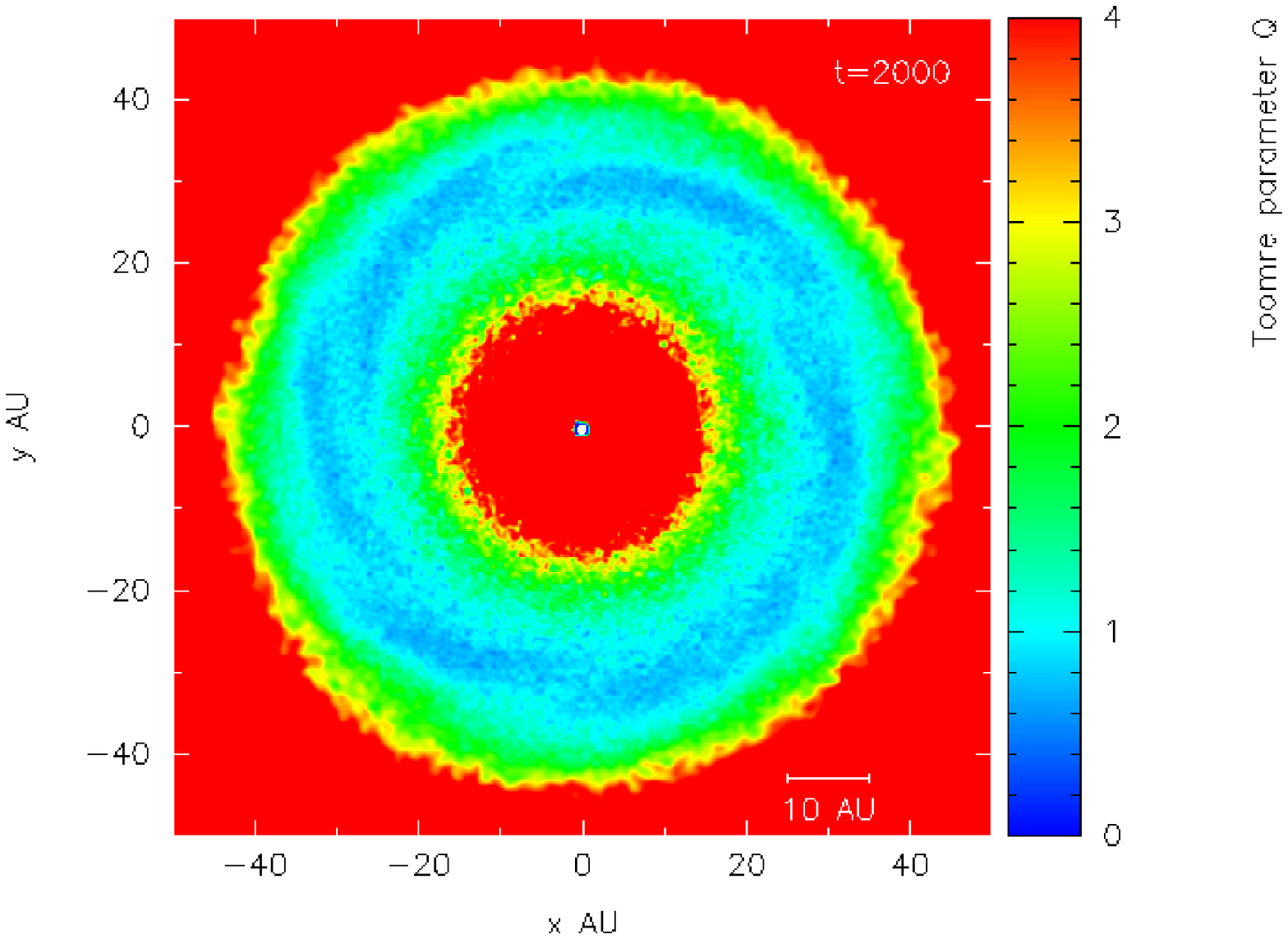}}
\caption{Toomre parameter for a disc irradiated by a $10\,{\rm K}$ background radiation field.}
\label{fig:dur4b.toomre}
\centering{
\includegraphics[height=9cm,angle=-90]{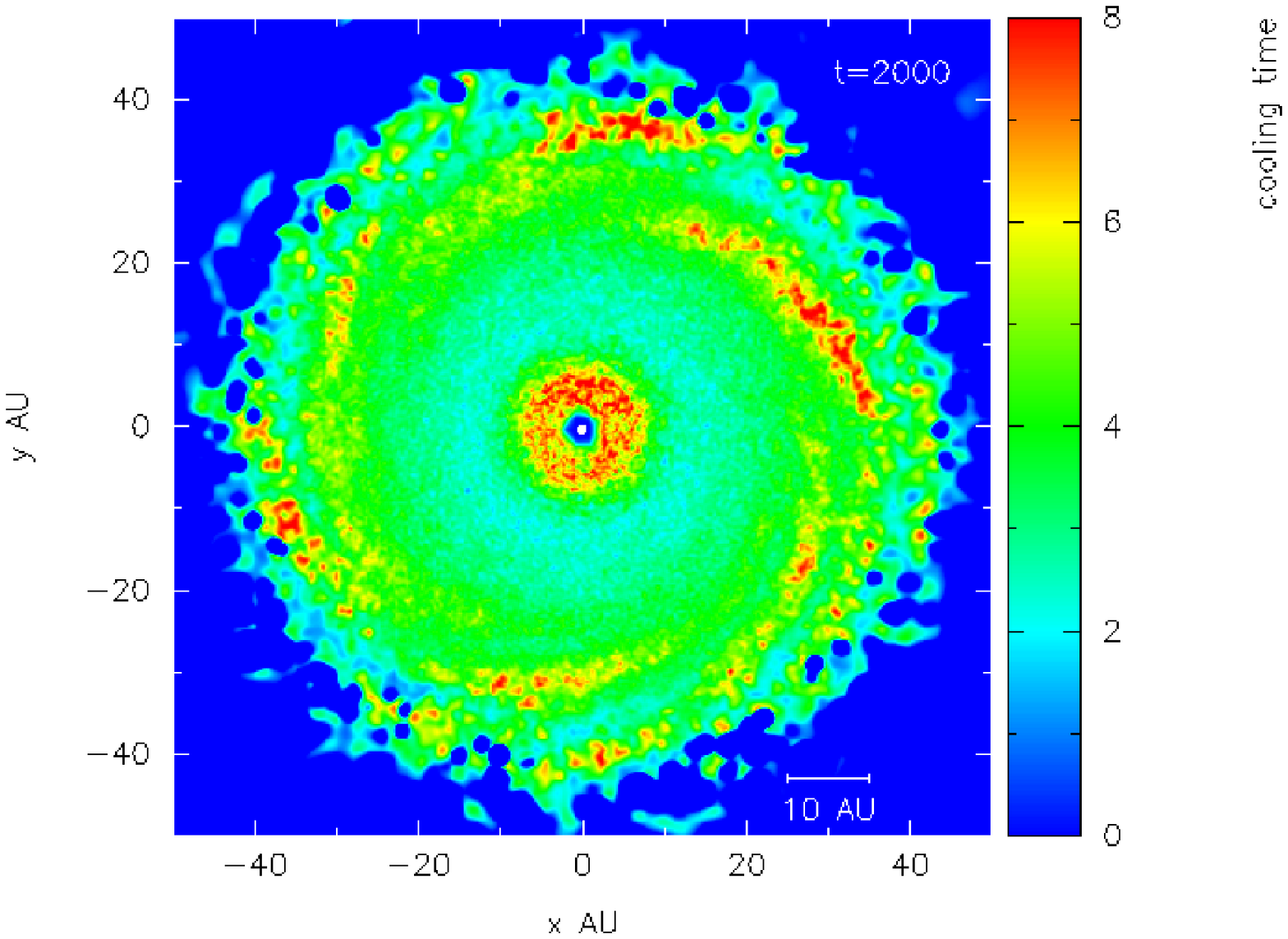}}
\caption{Net cooling time (in units of the local orbital period, and averaged vertically) for a disc irradiated by a $10\,{\rm K}$ background radiation field.}
\label{fig:dur4b.cool}
\end{figure}

\subsection{Disc evolution in a $\,10\,{\rm K}\,$ blackbody background radiation field}
\label{disc1}

In this case, the disc relaxes to a quasi-steady state within $\sim 250$~yr; this is the {\it asymptotic phase} defined by Boley et al. (2006). Thereafter, the disc cools slowly. In Figs.~\ref{fig:dur4b.dens}-~\ref{fig:dur4b.toomre} we plot the surface density of the disc, the midplane temperature, the Toomre parameter $Q$, and the net cooling time, at $t=2000$~yr. Weak spiral arms form in the disc but they show no tendency to fragment (see Fig.~\ref{fig:dur4b.dens}), despite the fact that the disc is Toomre unstable at $\sim 30$~AU (Fig.~\ref{fig:dur4b.toomre}). This is because the disc cannot cool fast enough; the net cooling time throughout the disc is generally $\stackrel{>}{_\sim}2 t_{_{\rm ORB}}$ (Fig.~\ref{fig:dur4b.cool}), where  $t_{_{\rm ORB}}$ is the local orbital period.\footnote{In the inner parts of the disc, the ratio of specific heats decreases from $\gamma\simeq 5/3$ to $\gamma\simeq 7/5$, due to the increasing temperature and the resulting excitation of the rotational degrees of freedom of H$_{_2}$. Consequently the maximum value of $t_{_{\rm COOL}}$ for fragmentation increases to $\sim 2t_{_{\rm ORB}}$ (e.g. Johnson \& Gammie 2003; Rice et al. 2005).}

\begin{figure}
\centering{
\includegraphics[height=6.9cm,angle=-90]{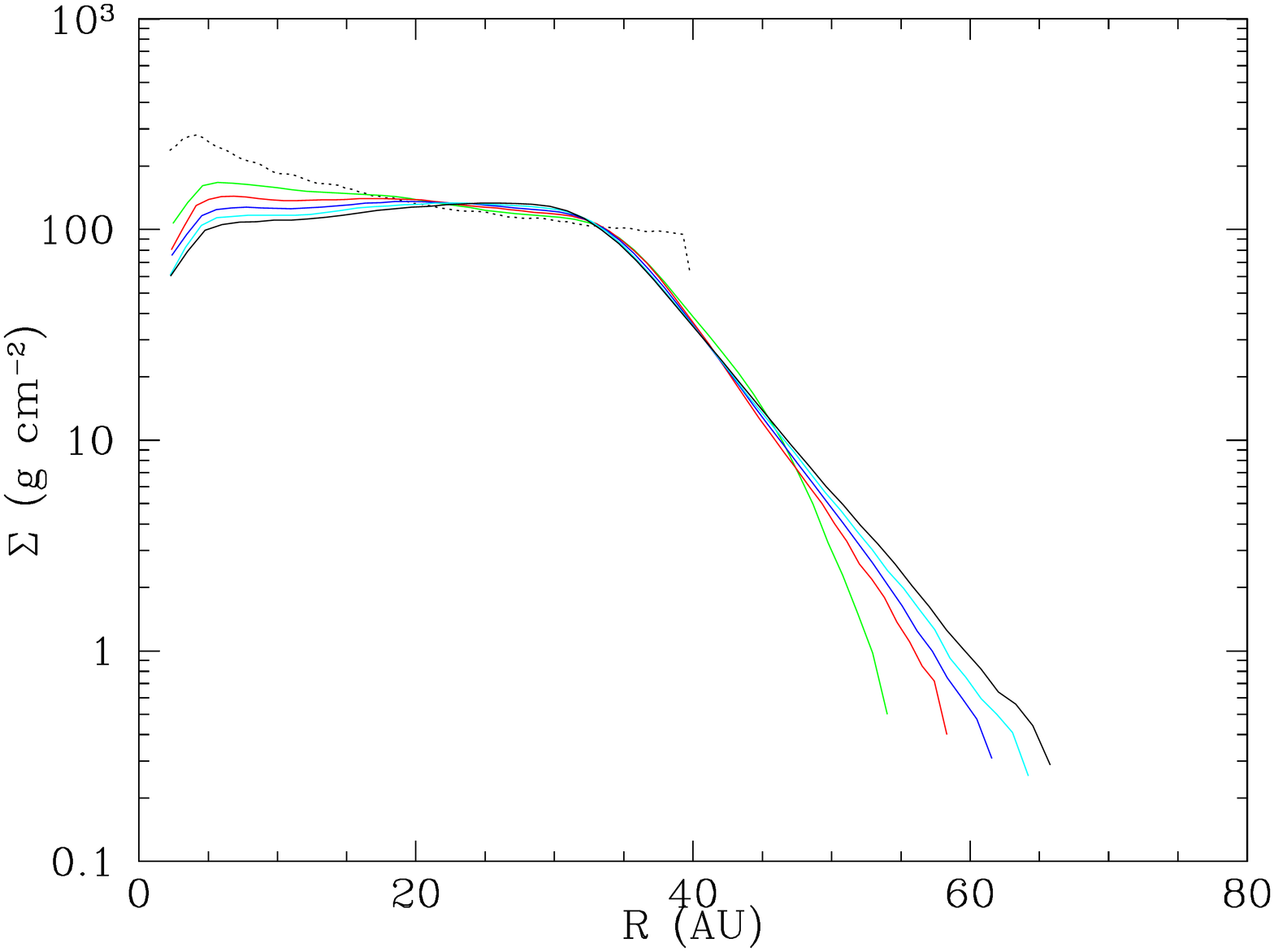}
\includegraphics[height=6.9cm,angle=-90]{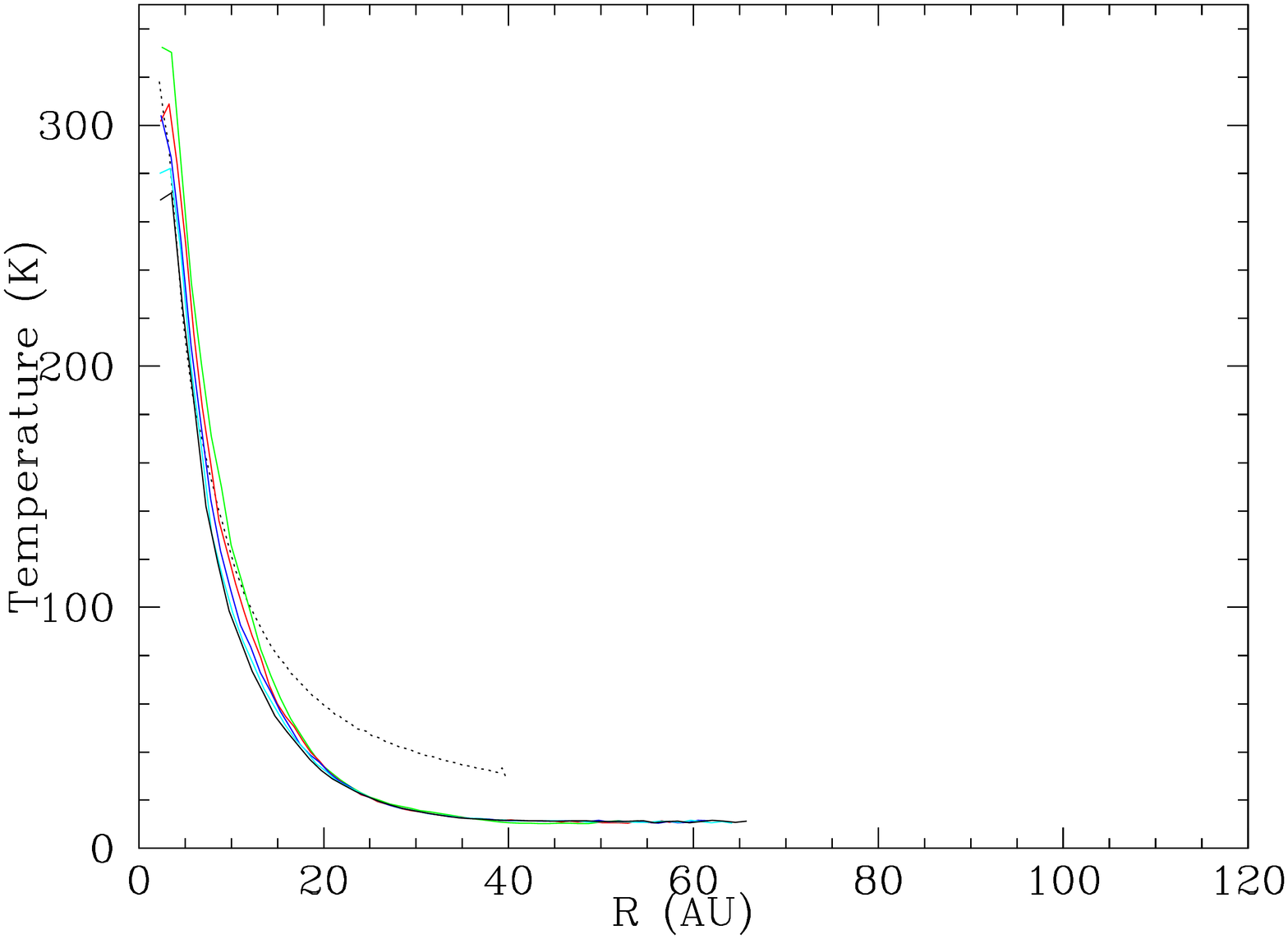}}
\caption{Surface density (azimuthally averaged) and temperature (azimuthally and vertically averaged) at five times ($t=500,1000,\dots,2500$~yr; green, red, blue, cyan, black) during the evolution of a disc irradiated by a $10\,{\rm K}$ background radiation field. The dotted lines correspond to the initial conditions.}
\label{fig:dur4b.qst}
\end{figure}

\begin{figure}
\centering{
\includegraphics[height=6.9cm,angle=-90]{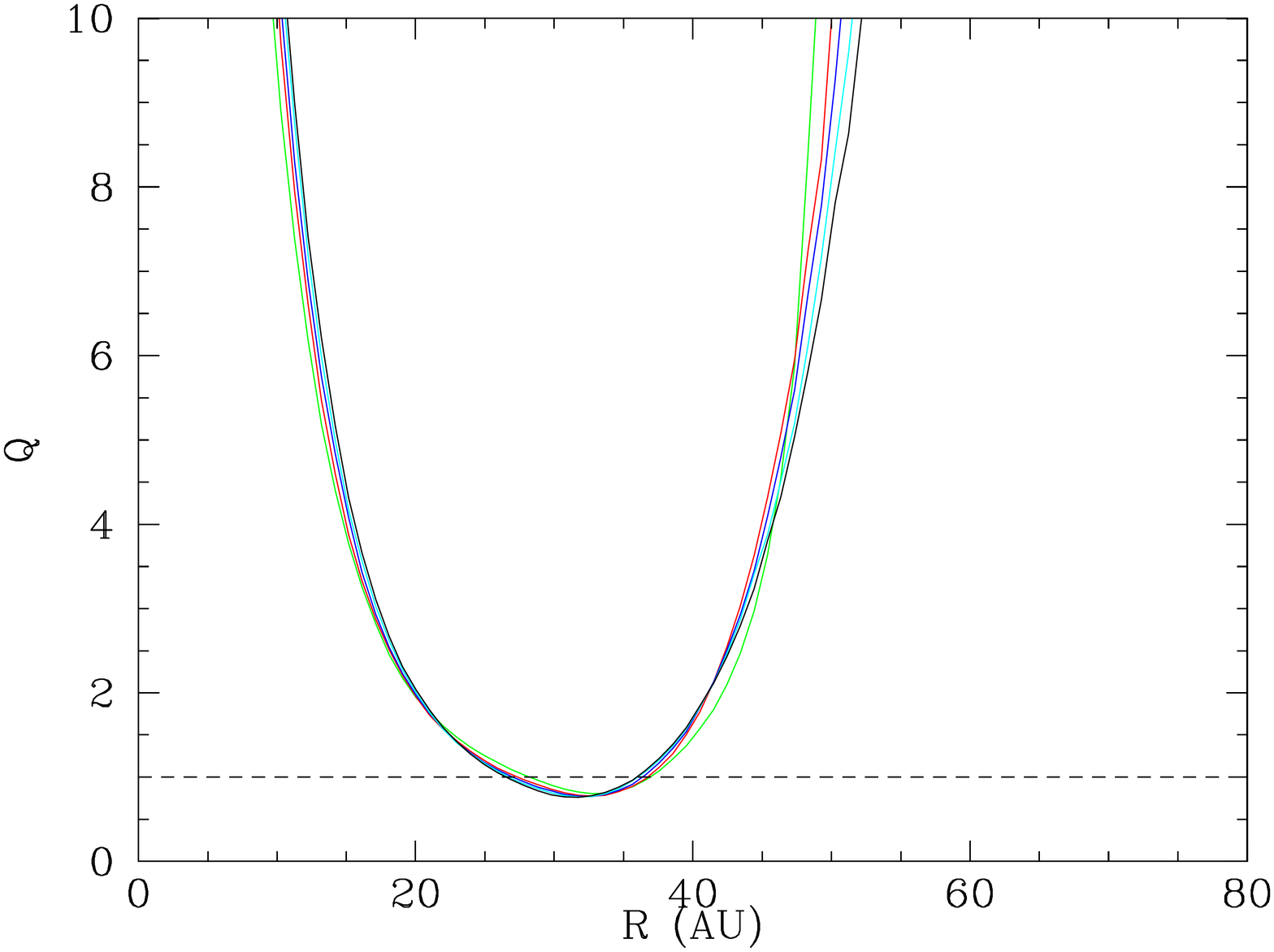}
\includegraphics[height=6.9cm,angle=-90]{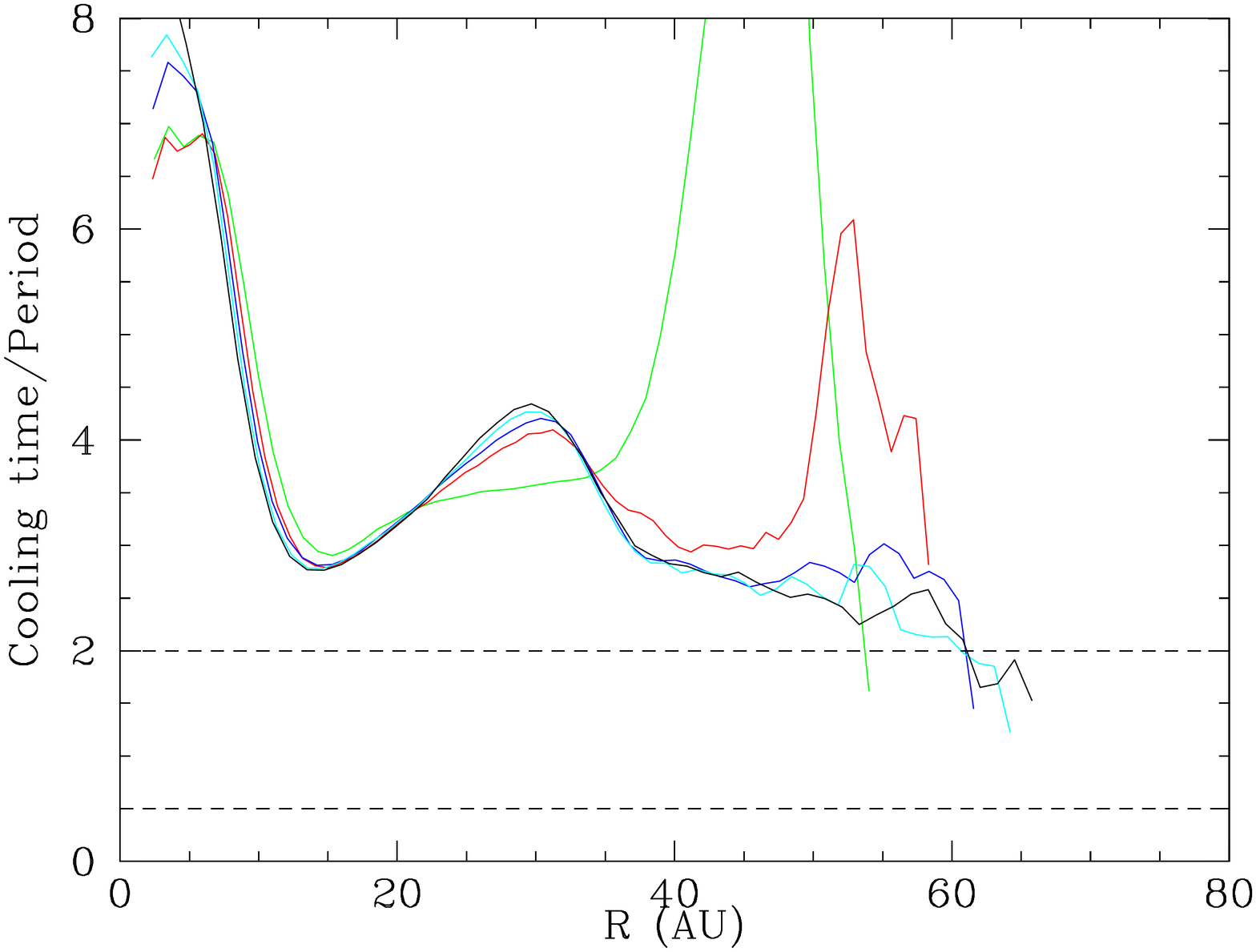}}
\caption{As, Fig.~\ref{fig:dur4b.qst}, the Toomre parameter (azimuthally averaged) and net cooling time (in units of the local orbital period, also azimuthally averaged).}
\label{fig:dur4b.qtc}
\end{figure}

In Figs.~\ref{fig:dur4b.qst} and \ref{fig:dur4b.qtc} we plot the azimuthally averaged surface density, temperature, Toomre parameter and cooling time (in units of the local orbital period) at five times during the disc evolution ($t=500,1000,\dots,2500$~yr). The temperature and cooling time are also averaged vertically relative to the disc midplane, and the Toomre parameter is calculated using the midplane isothermal sound speed. The profiles are essentially independent of time, i.e. the disc is in a quasi-steady state.
 
This simulation was repeated using a smaller number of SPH particles ($5\times10^4$) and the evolution of the thermal energy, surface density, temperature, Toomre parameter and cooling time were essentially unchanged, indicating that the simulation is converged.


\subsubsection{Comparison with the simulations of Boley et al. (2006) and Cai et al. (2007)}

The results of the simulations presented here are similar to those of Boley et al. (2006) and Cai et al. (2007). In this subsection, we make a detailed comparison.

(i) The temperatures we obtain are generally larger than those obtained by Boley et al. (2006), on average by $\sim 15\%$. At small radii our higher temperatures are probably due to higher viscous heating. At large radii ($\stackrel{>}{_\sim}30$~AU) the temperature differences are due to using different background radiation fields:in our simulation $T_{_{\rm BGR}}=10$~K, whereas in the Boley et al. simulation the disc can cool down to 3~K.

(ii) In the quasi-steady state, the Toomre parameter in our simulation is {\it generally} higher than that of Boley et al. (2006). This is because in the Boley et al. simulation the temperatures are lower, and -- due to the strong burst phase, i.e. a phase where violent gravitational instabilities occur and rapidly redistribute angular momentum within the disc -- the surface densities are higher. However, the region around $30\,{\rm AU}$ does not follow this trend, with the Toomre parameter in our simulations being smaller ($\sim\!0.8$ compared with $\sim\!1.45$ in Boley et al.); this is due to our disc having higher surface density at this radius.\footnote{Note also that Boley et al. use the adiabatic sound speed to calculate Q, and hence their $Q$-values are increased by a factor $\gamma^{1/2}\sim 1.3$ relative to ours.}

(iii) Our disc relaxes rapidly to its equilibrium state, within one disc rotation, without the `burst'  phase reported by Boley et al. (2006). We attribute this to two effects. Firstly, the higher temperatures in our disc damp gravitiational instabilities more rapidly. Secondly, the larger initial density fluctuations ($\propto N^{-1/2}\sim 0.0014$, where $N$ is the number of SPH particles) allow angular momentum to be redistributed by gravitational torques without first having to wait for substructure to develop by gravitational instabilities. In the Boley et al. simulations, the Toomre parameter becomes much lower, and so more violent gravitational instabilities develop, and the resulting spiral arms re-distribute angular momentum. Cai et al. (2007) and Boley (private communication) confirm that more powerful ambient heating and/or greater noise tend to weaken the burst phase.

(iv) In the quasi-steady state, our cooling times are similar to those of Boley et al. in the region around $30\,{\rm AU}$, but somewhat lower outside this region. These differences are attributable to the differences in surface density and temperature noted above.

\begin{figure}
\centering
\includegraphics[width=9cm]{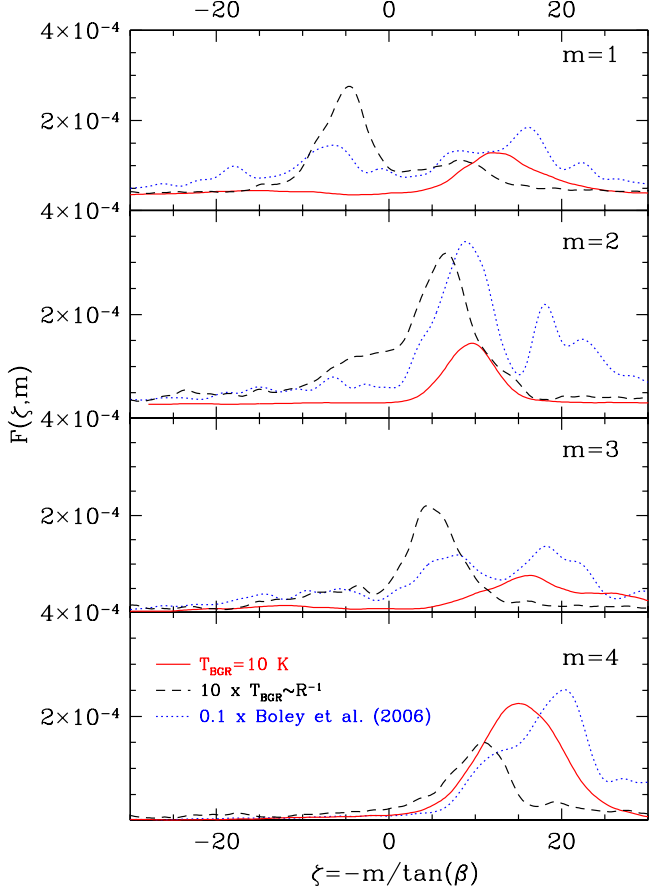}
\caption{The mean strengths, $F$, of the $m=1,2,3$ and 4 spiral modes during the quasi-steady phase, against $\zeta=-m/\tan(\beta)$, where $\beta$ is the pitch angle, for a disc irradiated by a $10\,{\rm K}$ blackbody background radiation field (solid lines), a disc irradiated by the central star ($\times 10$; dashed lines), and the Boley et al. (2006) disc which is irradiated by a $3\,{\rm K}$ blackbody background radiation field ($\times 0.1$; dotted lines).}
\label{fig:modes}
\end{figure}

(v) The spiral arms that develop in our disc are weaker than those reported by Boley et al. (2006).  To quantify this difference, we decompose the disc structure into a sum of Fourier components. We use as our basis a logarithmic spiral, $R=R_0 e^{-m \phi /\zeta}$, where $m$ is the mode of the perturbation, $\phi$ is the azimuthal angle of the SPH particle, and $\zeta=-m/\tan(\beta)$ is a parameter that represents the pitch angle $\beta$ of the spiral (Sleath \& Alexander 1996). The Fourier transform is then
\begin{eqnarray}\nonumber\label{eq:fourier}
F(\zeta,m)&=&\int_{-\infty}^{\infty}\int_{-\pi}^{\pi}\sum_{j=1}^{N}\left\{\delta\left(u\!-\!\ln[R_j]\right)\delta(\phi\!-\!\phi_j)\right\}e^{-i(\zeta u+m\phi)}dud\phi \\
 &= &\frac{1}{N}\sum_{j=1}^{N}{e^{-i\left(\zeta\ln[R_j]+m\phi_j\right)}},
\end{eqnarray}
where $(R_j,\phi_j)$ are the co-ordinates of particle $j$. In Fig.~\ref{fig:modes} we plot $F(\zeta,m)$ against $\zeta$ for the $m=1,2,3$ and 4 modes, for our simulation and for the Boley et al. (2006) simulation. For the Boley et al. data the sum in Eq.~(\ref{eq:fourier}) is over all points of their cylindrical grid, weighted according to the mass $m_j$ of each grid cell, i.e. $F(\zeta,m)=\sum_{j=1}^{N}{m_j\ e^{-i\left(\zeta\ln[R_j]+m\phi_j\right)}/\sum_{j=1}^{N}{m_j}}$.  The maxima in $F(\zeta,m)$ identify the dominant pitch angles of the spirals. In both cases, $F$ has been averaged over $~\sim 200$ yr during the quasi-steady state, and in the Boley et al. case $F$ has also been divided by 10 to make comparison easier. The spiral arms in the Boley et al. disc are about six times stronger than in our disc. 

(vi) In our simulation, the central star is represented by an accreting sink particle with radius $R_{_{\rm SINK}}=2\,{\rm AU}$, which keeps the density low near this radius. Consequently we do not see the dense rings that Boley et al. (2006) report in the inner regions of their disc.

(vii) In both simulations there is no tendency to fragment.

Our simulation is also comparable with the simulations by Cai et al. (2007) using background radiation fields of $15\,{\rm K}$ and $25\,{\rm K}$. We estimate that across the disc our temperatures are less than $3\%$ higher than in their $15\,{\rm K}$ case, and approximately the same as in their $25\,{\rm K}$ case. Their discs also show no tendency to fragment. We conclude that the Indiana University Hydrodynamics Group code (e.g. Pickett et al. 1998; Mejia et al. 2005) and our SPH-RT code produce very similar results. This is significant, because the treatments of radiative transfer and hydrodynamics are completely different in the two codes.

\subsubsection{The vertical temperature profile of the disc}

Our radiative hydrodynamic computational scheme has already been tested for the case of collapsing clouds and shown to perform well (Stamatellos et al. 2007). Here, we compare our results with the analytic prescription of Hubeny (1990) (cf. Boley et al. 2007b).

Hubeny (1990) has calculated the thermal structure of a cylindrically symmetric, stationary Keplerian disc. Radiation transport is allowed only in the vertical direction, i.e. perpendicular to the disc mid-plane. Energy is deposited by the shear of the Keplerian motion, with a viscosity which is independent of $z$. Self-gravity, convection and external irradiation are all neglected. At any radius, the temperature $T$ at optical depth $\tau$ from the surface of the disc then approximates to
\begin{equation}
\label{eq:hubeny}
T(\tau)=T_{\rm eff}\left\{\frac{3}{4}\left[(\tau-\tau{_\theta})+\frac{1}{\sqrt{3}}+\frac{2}{3\kappa_{_{\rm R}}(\tau){\Sigma}}\right]\right\}^{1/4},
\end{equation}
where $\sigma_{_{\rm SB}} T^4_{\rm eff}$ is the vertical flux at the disc surface, $\tau{_\theta}=2\int_{0}^{\sigma}{\kappa_{_{\rm R}} (\sigma^\prime)\ \sigma^\prime{\rm d}\sigma^\prime}/\Sigma$, and $\sigma(z)=\int_{z}^{\infty}\rho\ {\rm d}z$. Eqn. (\ref{eq:hubeny}) is derived from Hubeny's Eqn. (8)  by setting $\gamma_J=\gamma_H=1$ and $w(m)=\bar{w}$. 

\begin{figure}
\centering{
\includegraphics[width=7.2cm]{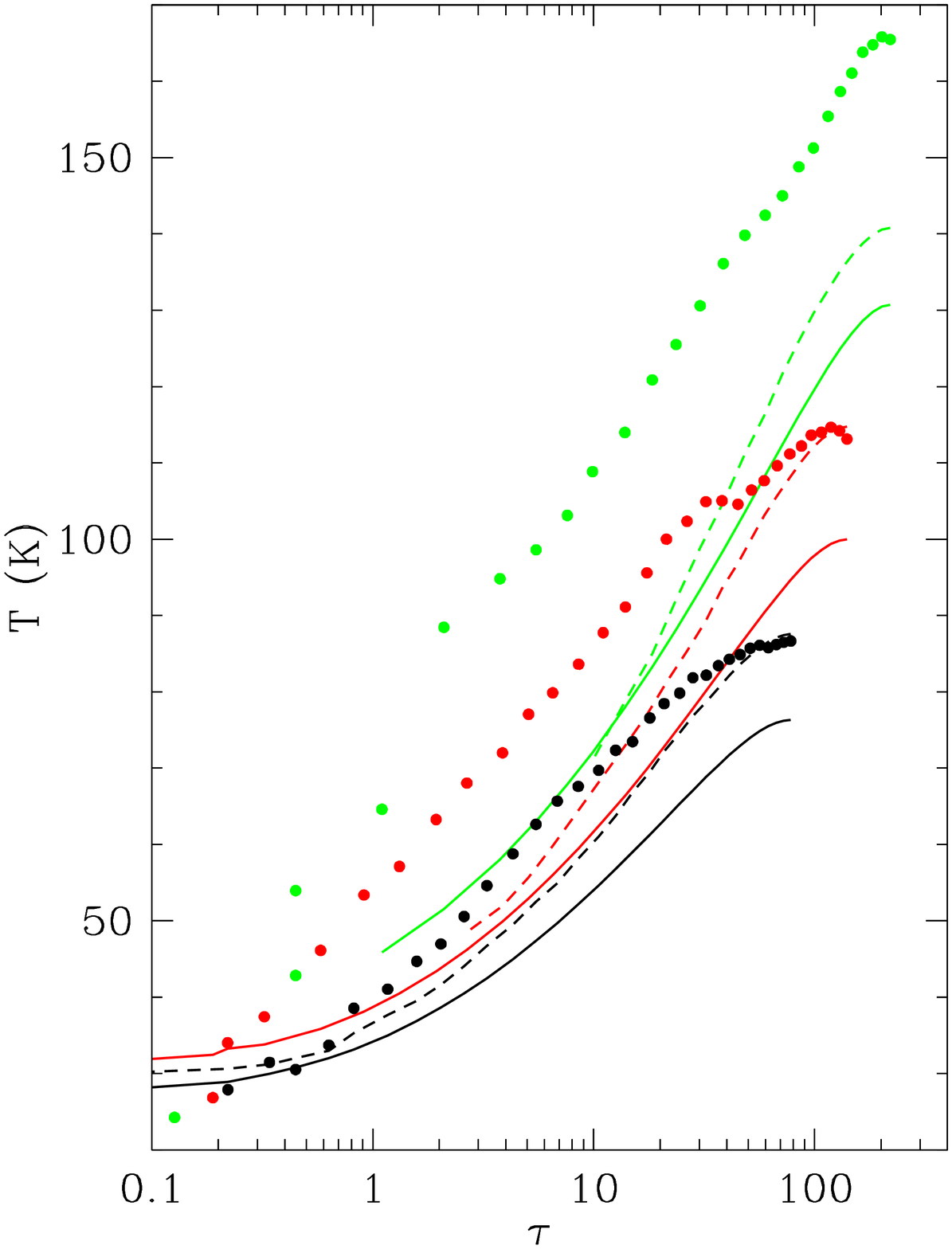}}
\centering{
\includegraphics[width=7.2cm]{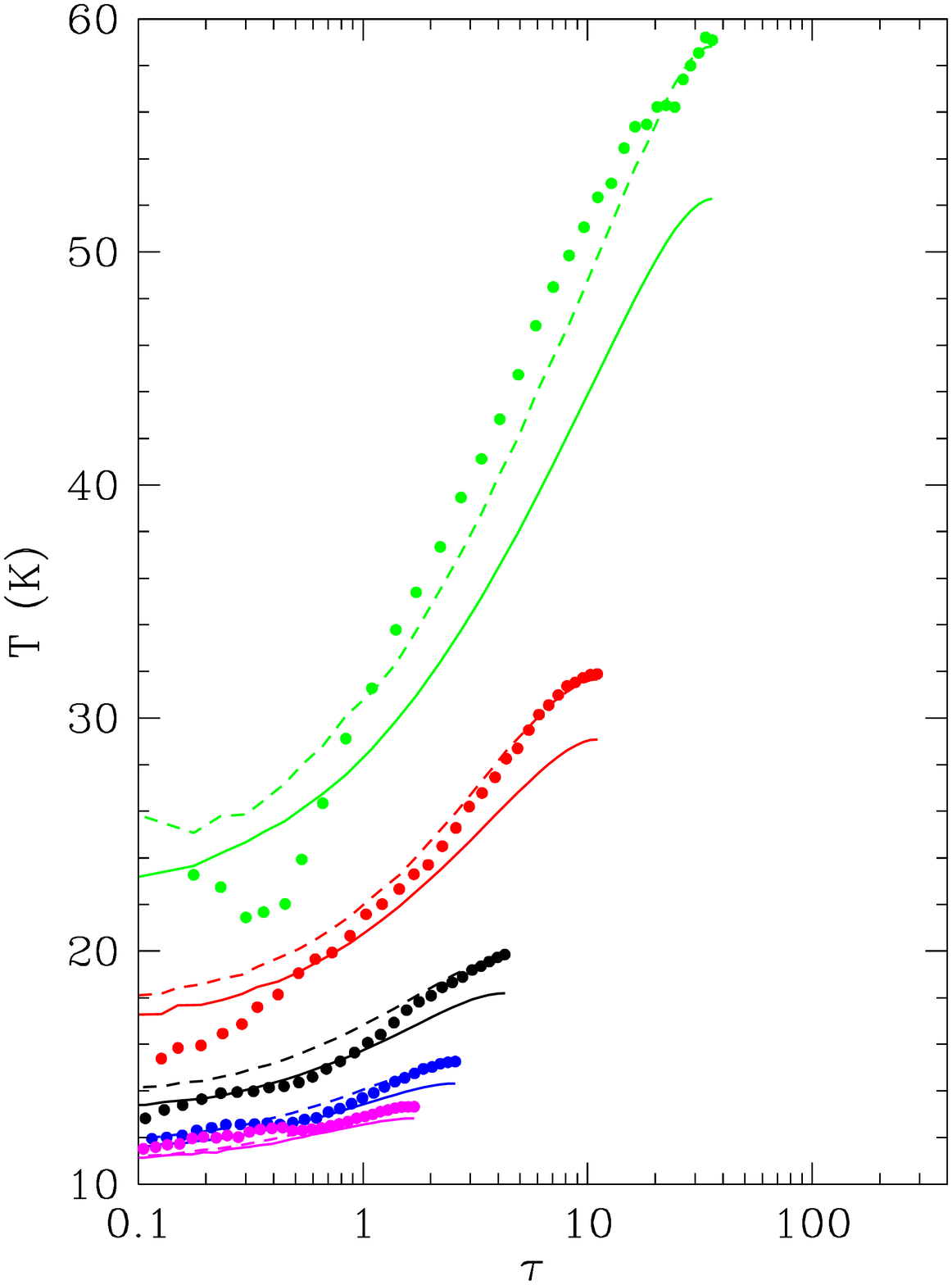}
}
\caption{Vertical temperature profiles, $T(\tau)$, at different radii (6, 9, 11~AU, top to bottom, top plot; 14, 20, 25, 30, 33~AU, top to bottom, bottom plot). Points correspond to the results of our model (azimuthally averaged), the solid lines to the Hubeny (1990) solution, and the dashed lines to the Hubeny solution corrected for the fact that radiation diffuses both vertically and radially.}
\label{fig:hubeny}
\end{figure}

We compare the Hubeny solution with our disc at $t=2000$ yr, during the quasi-steady phase. Our simulation accounts for external irradiation and for self-gravity, which are not included in the Hubeny solution; there are also weak spiral arms in our disk. In Fig.~\ref{fig:hubeny} the dots represent the vertical temperature structure in our disc at different radii (6, 9, $11\,{\rm AU}$, top plot; 14, 20, 25, 30, $33\,{\rm AU}$, bottom plot); here, temperature is plotted as a function of the vertical optical depth from the surface of the disc, $T(\tau)$. The solid lines represent the Hubeny solution at the same radii, when $T_{\rm eff}$ is estimated by summing the radiative losses $L$ from a thin annulus with cross-sectional area $A$ at each radius, and then putting $L=2\,A\,\sigma_{\rm SB}\, T_{\rm eff}^4$; the resulting Hubeny temperatures are lower than in our simulation, especially near the central star. However, the Hubeny solution assumes plane-parallel symmetry, with a purely vertical temperature gradient. In our disc there is also a significant radial temperature gradient (see Fig.~\ref{fig:dur4b.qst}), and therefore energy diffuses both vertically and radially.This results in temperatures near the midplane that are higher than the Hubeny solution predicts. To correct the Hubeny solution for this effect,  we calculate the vertical optical depth, $\tau$, to the disc surface at each position of the disc, and assume that this optical depth also represents how far the radiation has diffused radially in the disc. We then assume that the temperature given by the Hubeny solution at $(\tau,\tau_R-\tau)$, where $\tau_R$ is the radial optical depth from the inner edge of the disc,  actually represents the disc temperature at $(\tau,\tau_R)$. The corrected Hubeny temperature profiles (dashed lines) are better fits to our disc, apart from the inner inner parts ($<8$~AU), where our temperatures remain higher because they reflect additional heating sources that are not included in the Hubeny solution.

\begin{figure}
\centering{
\includegraphics[height=9cm,angle=-90]{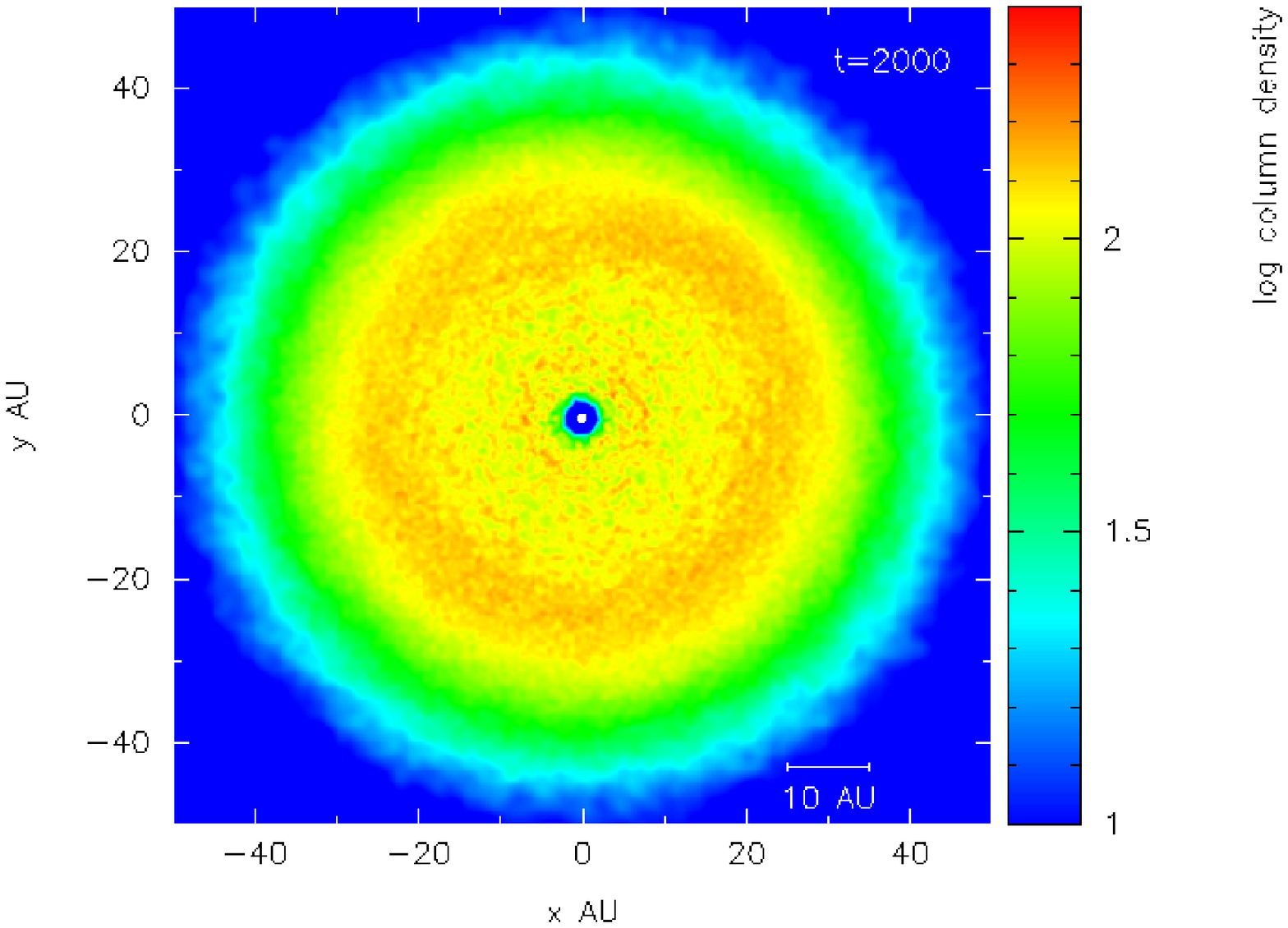}
\includegraphics[height=8.3cm,angle=-90]{dens.xz.4a.ps}}
\caption{As Fig. \ref{fig:dur4b.dens}, but taking account of radiation from the central star.}
\label{fig:dur4a.dens}
\centering{
\includegraphics[height=9cm,angle=-90]{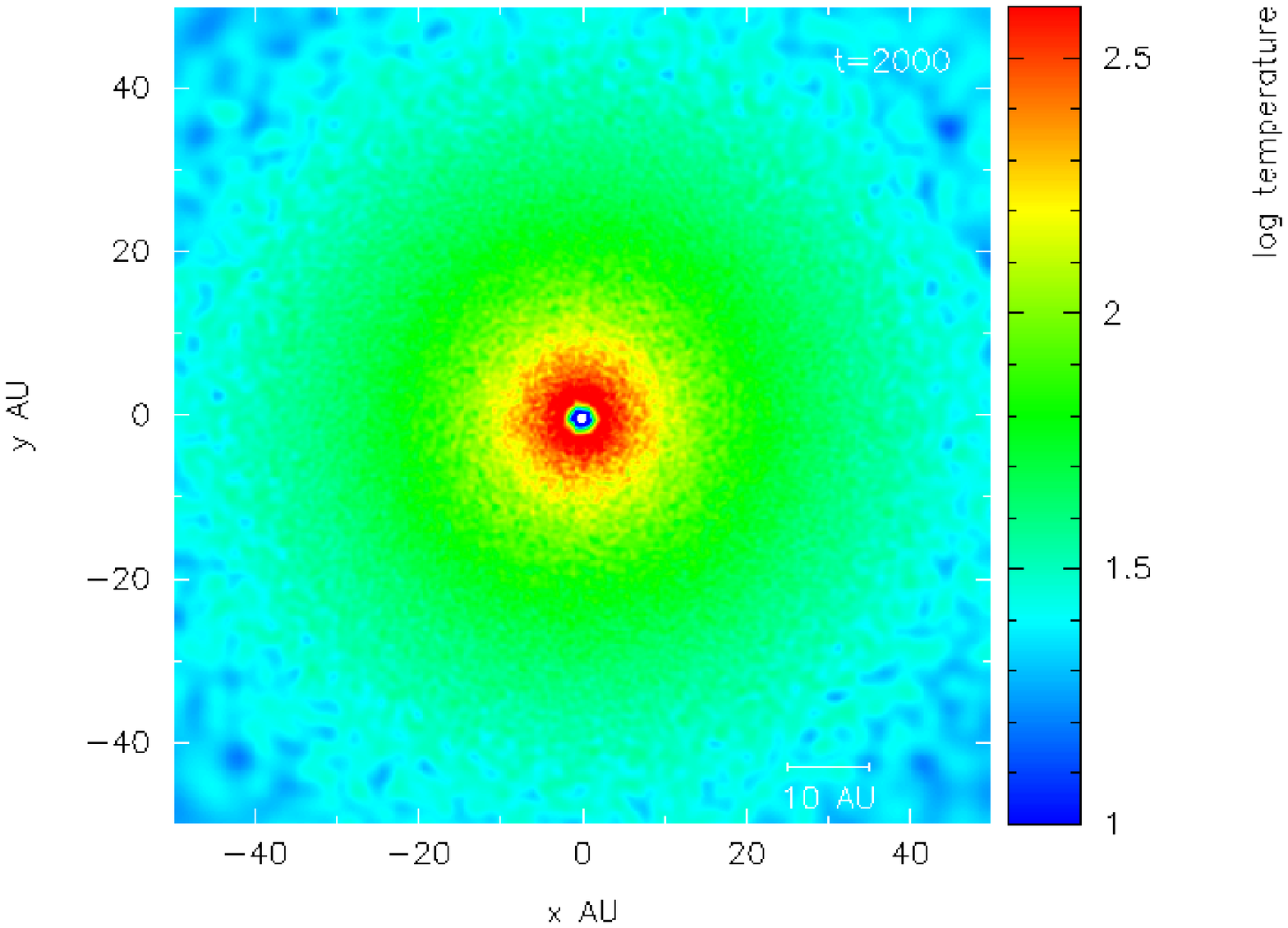}
\includegraphics[height=8.3cm,angle=-90]{temp.xz.4a.ps}}
\caption{As Fig. \ref{fig:dur4b.temp}, but taking account of radiation from the central star.}
\label{fig:dur4a.temp}
\end{figure}

\begin{figure}
\centering{
\includegraphics[height=9cm,angle=-90]{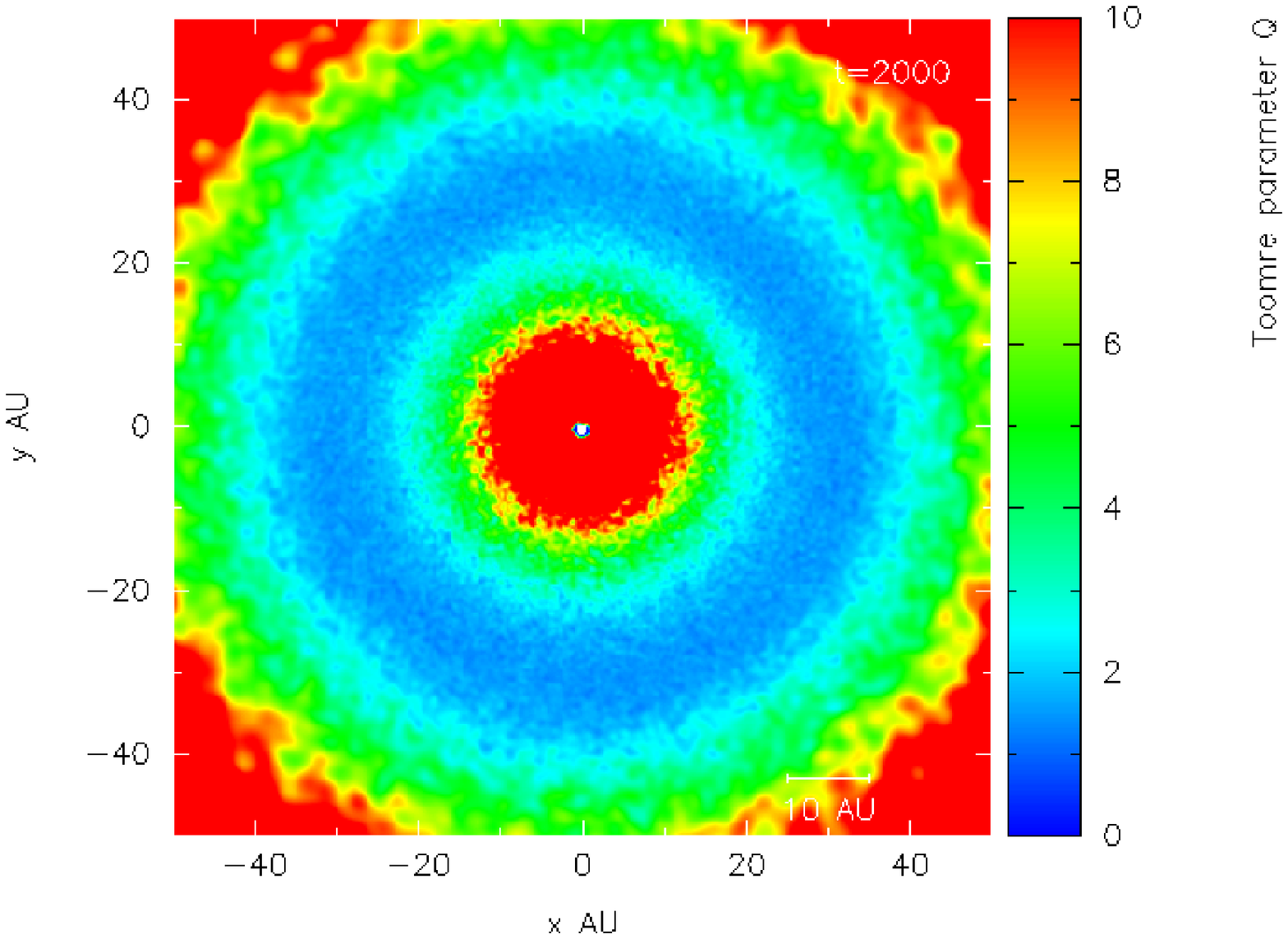}}
\caption{As Fig. \ref{fig:dur4b.toomre}, but taking account of radiation from the central star.}
\label{fig:dur4a.toomre}
\centering{
\includegraphics[height=9cm,angle=-90]{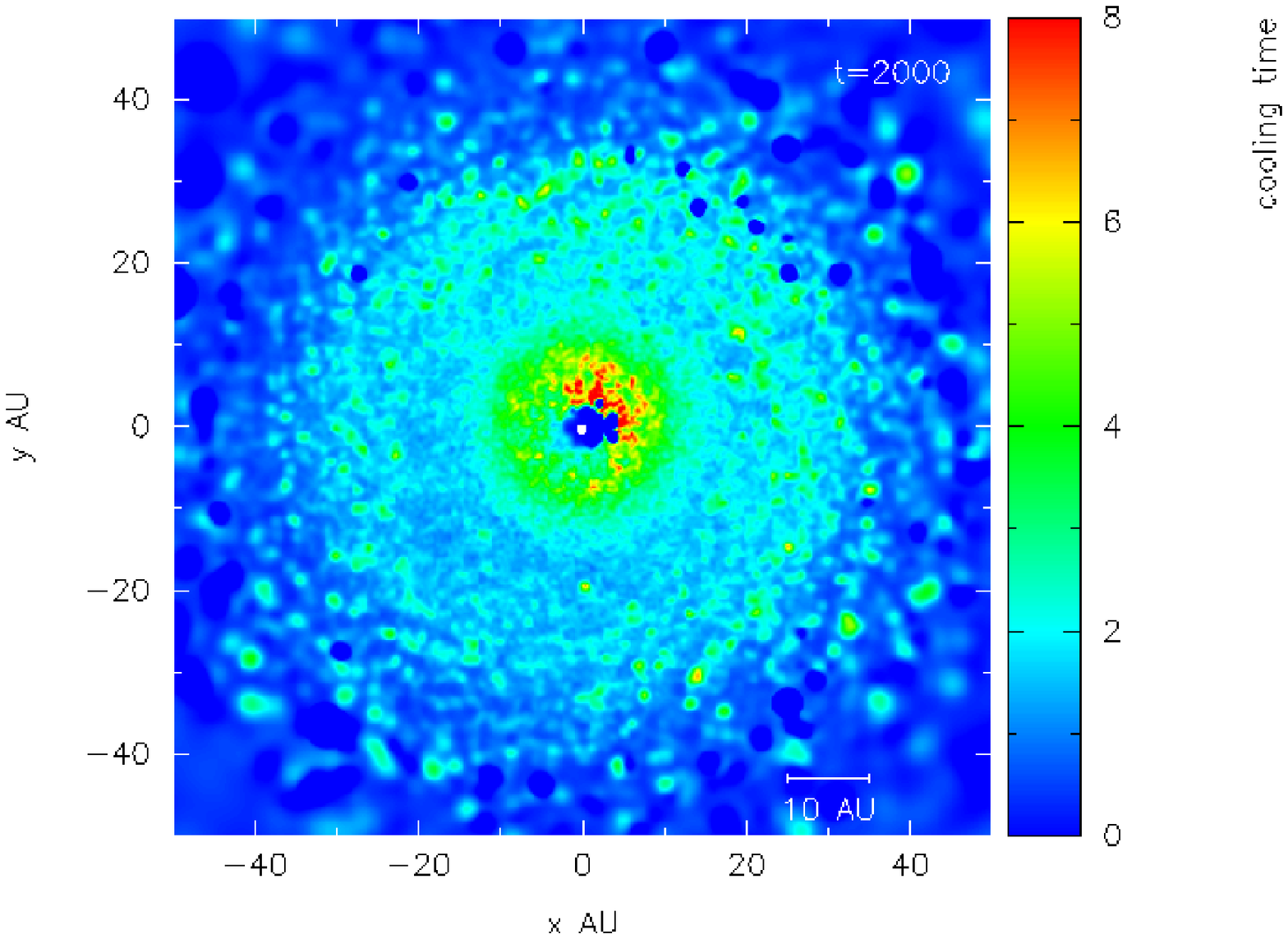}}
\caption{As Fig. \ref{fig:dur4b.cool}, but taking account of radiation from the central star.}
\label{fig:dur4a.cool}
\end{figure}

\begin{figure}
\centering{
\includegraphics[height=6.9cm,angle=-90]{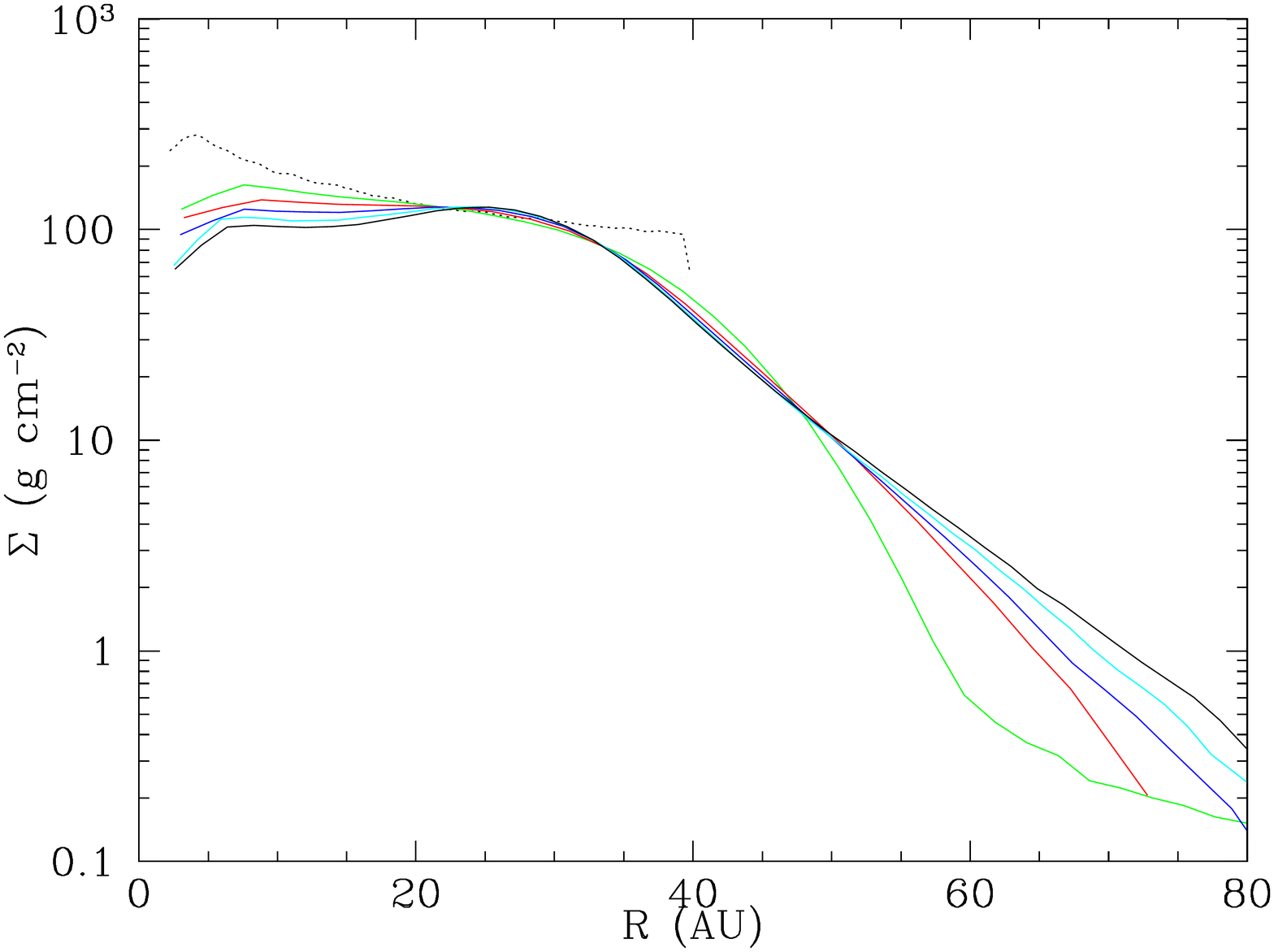}
\includegraphics[height=6.9cm,angle=-90]{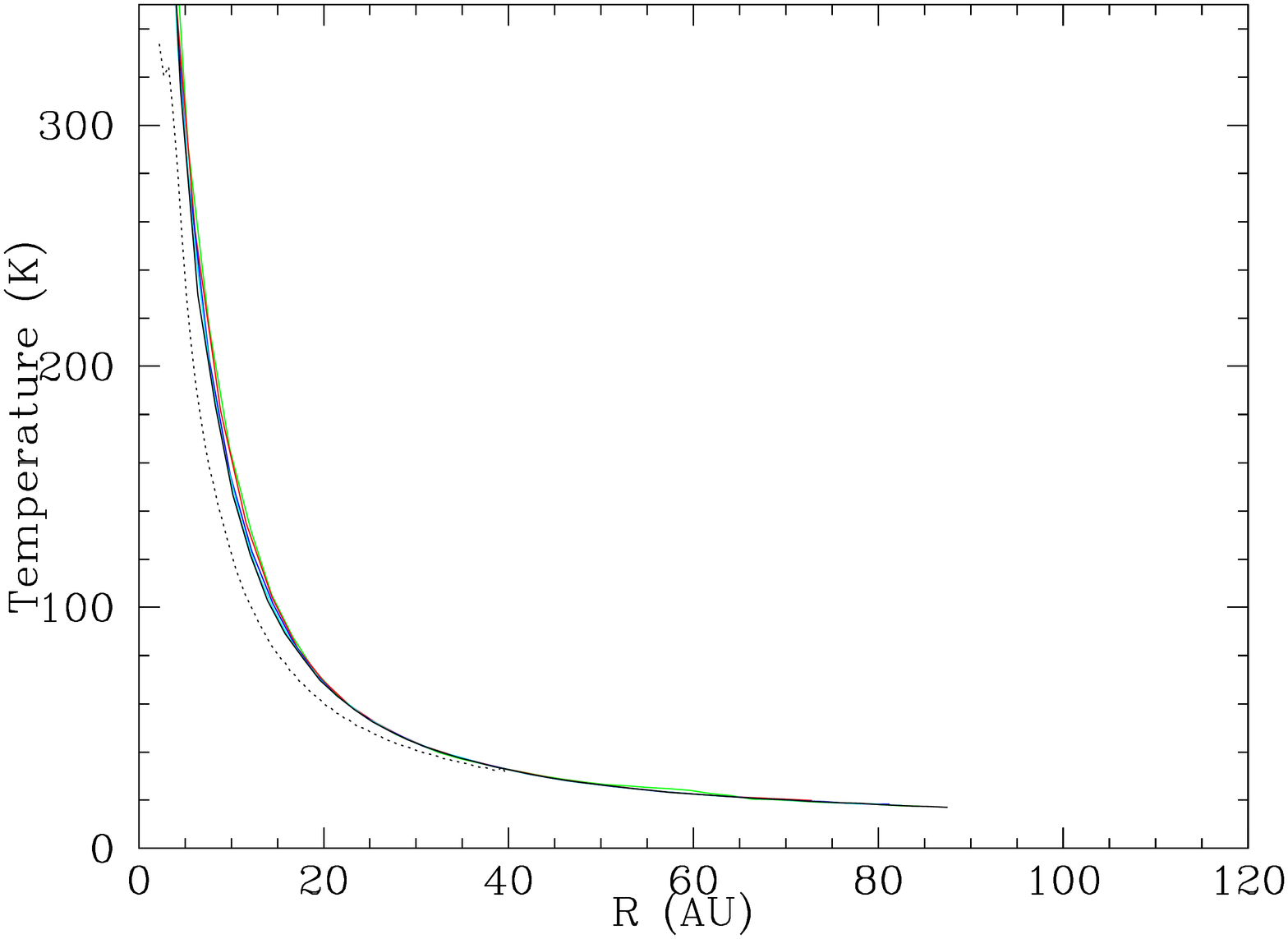}}
\caption{As Fig. \ref{fig:dur4b.qst}, but taking account of radiation from the central star.}
\label{fig:dur4a.qst}
\end{figure}
\begin{figure}
\centering{
\includegraphics[height=6.9cm,angle=-90]{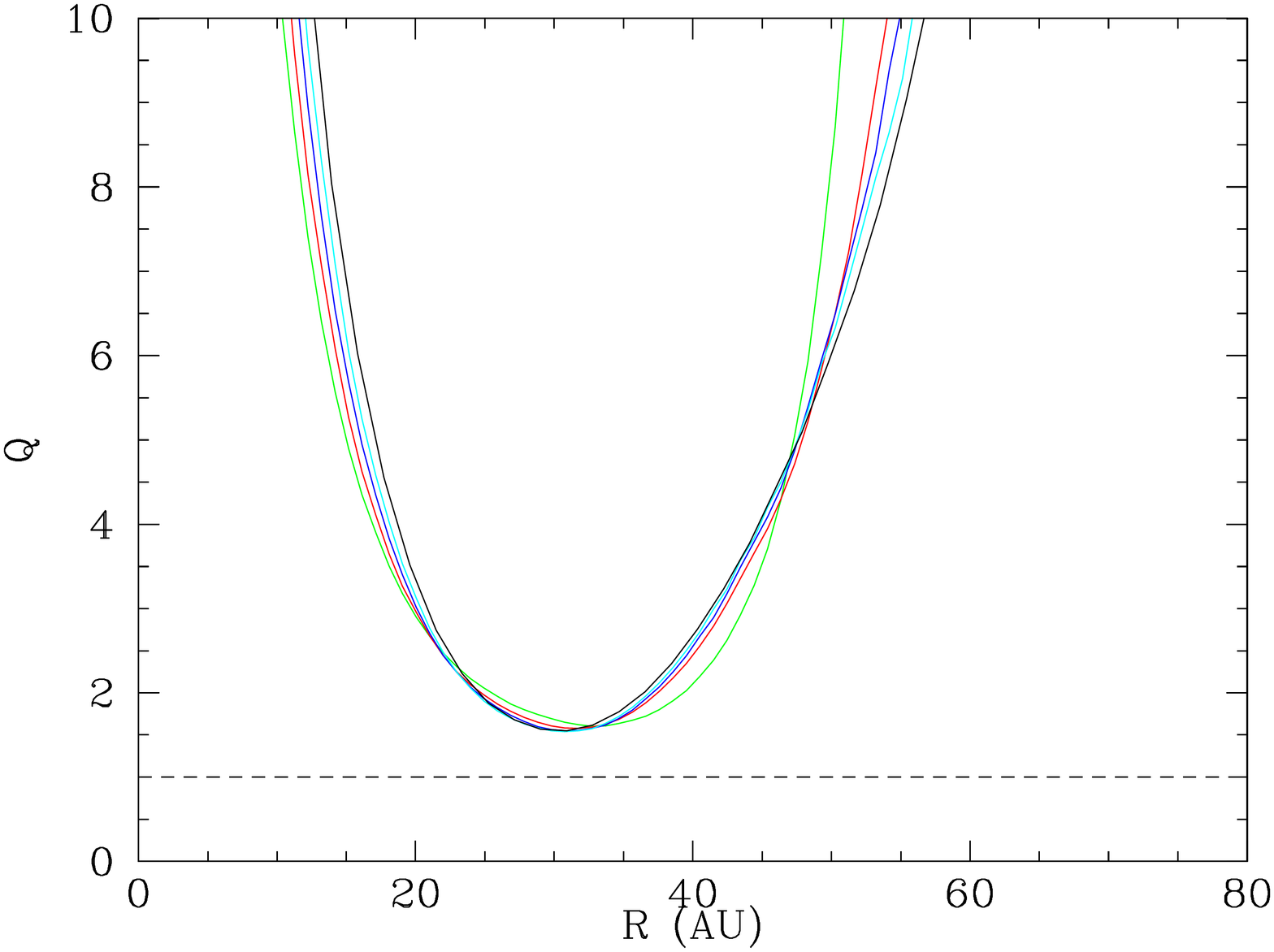}
\includegraphics[height=6.9cm,angle=-90]{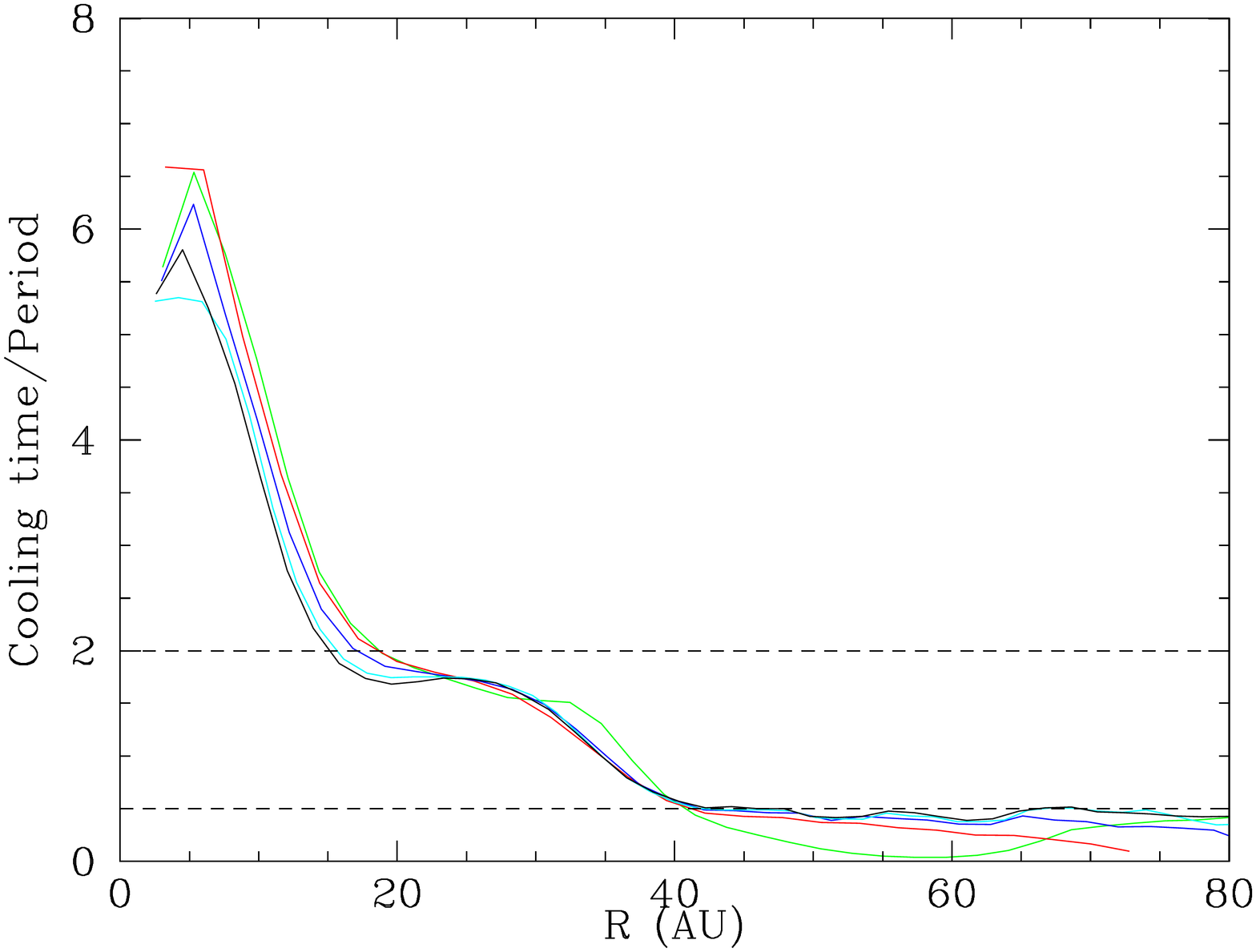}}
\caption{As Fig. \ref{fig:dur4b.qtc}, but taking account of radiation from the central star.}
\label{fig:dur4a.qtc}
\end{figure}

\subsection{Disc evolution taking account of radiation from the central star}
\label{disc2}

In reality the disc is likely to be heated by radiation from the central star, and this will influence its evolution. In order to treat this radiation properly, it is necessary to solve a complicated transfer problem, which involves {\it both} radiation which arrives directly from the central star, {\it and} radiation which first interacts with the more diffuse envelope above and below the disc, and is then scattered, or absorbed and re-radiated, towards the disc. If a static dust distribution is specified, and radiative equilibrium is assumed, this problem can be solved using modern Monte Carlo methods (e.g.  Wood et al. 2002; Whitney et al. 2003; Stamatellos et al. 2005; Pinte et al. 2006; see review by Watson et al. 2007). However, if an evolving dust distribution is involved, a consistent solution for the dynamics and the radiation transport is only feasible -- with current computing resources -- if simplifying assumptions are made (e.g. Dullemond et al. 2007, and references therein).

Here we simply assume that radiation from the central star can be represented by a blackbody background radiation field whose temperature, $T_{_{\rm BGR}}$, decreases with distance, $R$, from the star. Observations indicate that the disc temperature drops radially as $R^{-q}$ with $0.4\leq q\leq 0.8$ (e.g. Beckwith et al. 1990; Osterloh \& Beckwith 1995). Hence, we set
\begin{equation}
T_{_{\rm BGR}}(R)=\left[T_0^2\left(\frac{R}{AU}\right)^{-2}+T_{\infty}^2\right]^{1/2}\,,
\end{equation}
in Eqn.~(\ref{eq:radcool}), with $T_0=1200$~K and $T_\infty=10$~K. We refer to this prescription as `implicit stellar irradiation'. (For simplicity, we adopt the same profile for the {\it initial} temperature of the matter in the disc, cf. Eqn.~\ref{eq:tempprofile}).

The results of this simulation are shown in Figs.~\ref{fig:dur4a.dens}-\ref{fig:dur4a.qtc}. In this case the the disc is initially close to equilibrium, and so its temperature hardly changes (see Fig.~\ref{fig:dur4a.qst}). Due to the implicit stellar irradiation, it is hotter -- than the disc irradiated by a $10\,{\rm K}$ background radiation field -- and cools fast enough to fragment at large radii $\stackrel{>}{_\sim}30\,{\rm AU}$ (Fig.~\ref{fig:dur4a.qtc}). However, it does not fragment because it is not gravitationally unstable ($Q\stackrel{>}{_\sim}1.5$; Fig.~\ref{fig:dur4a.qtc}).  

Implicit stellar irradiation stabilizes the disc. Fourier analysis reveals the presence of very weak spiral arms (see Fig.~\ref{fig:modes}), ten times weaker than for the disc irradiated by a $10\,{\rm K}$ background radiation field. This is consistent with simulations of discs heated by ``envelope-radiation" (e.g. Boss 2001, 2002; Cai et al. 2007), and with analytical predictions (Matzner \& Levin 2005; Rafikov 2005; Whitworth \& Stamatellos 2006).

Comparing our simulation with the $T_{_{\rm BGR}}\!=\!50$~K case of Cai et al. (2007), we find similar $Q$-values across their disc and ours. However, our disc is less extended than theirs, since they start this simulation from a late phase of their $T_{_{\rm BGR}}\!=\!25$~K case, i.e. after the disc has spread out. We estimate that the $m=1\;{\rm and}\;4$ modes are about five times stronger in their simulation than in ours (see Fig.~\ref{fig:modes}).

\section{Summary and conclusions}

We have performed radiative hydrodynamic simulations of the inner regions of circumstellar discs with parameters broadly similar to those used by Boley et al. (2006) and Cai et al. (2007), i.e. a $0.07~{\rm M}_{\sun}$ disc around a $0.5 ~{\rm M}_{\sun}$ star. Initially, the disc extends from 2 to 40~AU, with surface density $\Sigma\sim R^{-1/2}$, and temperature $T\sim R^{-1}$. 

We use a new method to treat the energy equation, which includes excitation of the rotational degrees of freedom of molecular hydrogen, and radiative transfer using realistic dust opacities.  Our simulations do not have sufficient resolution to capture convective energy transport. However, since proto-fragments must condense out on a dynamical scale, they do not have sufficient time to cool by convection, because this would require convective cells to migrate and disperse supersonically (Whitworth et al. 2007). Moreover, the surface densities of our discs never reach the high values which -- according to  Mayer \& Gawryszczak (2007) -- lead to proto-fragments cooled by convection. 

We have simulated the evolution of a disc irradiated by a cool background radiation field ($T_{_{\rm BGR}}=10\,{\rm K}$). Despite the fact that we use completely different treatments for the hydrodynamics and the radiative transport, our results are similar to those of Boley et al. (2006) and Cai et al. (2007). Our disc is gravitationally unstable at $\sim 30$~AU, and develops weak spiral arms, but it shows no tendency to fragment, because it cannot cool fast enough.

We have also simulated the evolution of a disc taking account of radiation from the central star ($T_{_{\rm BGR}}\sim R^{-1}$). In this case the disc can cool fast enough to fragment, because of its higher temperature, but it does not fragment, because it is not gravitationally unstable; it does not even develop noticeable spiral arms. This result agrees with previous simulations of discs heated by envelope radiation (e.g. Boss 2001, 2002; Cai et al. 2007). It also corroborates the analytic predictions of Matzner \& Levin (2005), Rafikov (2005) and Whitworth \& Stamatellos (2006).

Whitworth \& Stamatellos (2006) have argued that a massive disc {\it will} fragment at {\it larger} radii ($R\stackrel{>}{_\sim }100$~AU), producing brown dwarfs and occasionally low-mass hydrogen-burning stars or planetary-mass objects. These predictions are corroborated by the numerical simulations of Stamatellos, Hubber \& Whitworth (2007). However, if planetary mass objects form at such large radii, they are unlikely to migrate inwards to become hot Jupiters. More likely they are ejected into the field though 3-body interactions.

We conclude that observed gas giant planets in close orbits are unlikely to have formed by gravitational instability.

\begin{acknowledgements}
  
We would like to thank the referee R. Durisen for his comments that helped to improve the original manuscript. We also thank A. Boley  for providing data from Boley et al. (2006), and K. Rice for useful discussions. Colour plots were produced using {\sc splash} (Price 2007). The computations reported here were performed using the UK Astrophysical Fluids Facility (UKAFF).  We also acknowledge support by PPARC grant PP/E000967/1. 

\end{acknowledgements}

\end{document}